\documentclass[conference,compsoc]{IEEEtran}
\ifCLASSOPTIONcompsoc
  \usepackage[nocompress]{cite}
\else
  \usepackage{cite}
\fi

\ifCLASSINFOpdf
\else
\fi
\usepackage{amsmath}
\usepackage{pdfpages}
\usepackage{epstopdf}
\usepackage{pslatex}
\usepackage{subfig}
\usepackage{url}
\usepackage{pdfpages}
\usepackage{amsmath} %matrix
\usepackage{amssymb}
\usepackage{multicol}
\usepackage{lipsum}
\usepackage{lipsum}
\usepackage{mathtools}

\usepackage{cuted}

\usepackage{epsfig}
\usepackage{color}

\begin{document}

\title{Probabilistic Interpretation for Correntropy with Complex Data }

%\author{Jo\~ao P. F. Guimar\~aes, Aluisio I. R. Fontes, Joilson B. A. Rego, Allan de M. Martins}
\author{\IEEEauthorblockN{Jo\~ao P. F. Guimar\~aes}
\IEEEauthorblockA{joao.guimaraes@ifrn.edu.br}
\and
\IEEEauthorblockN{Aluisio I. R. Fontes}
\IEEEauthorblockA{aluisio.rego@ifrn.edu.br}
\and
\IEEEauthorblockN{Joilson B. A. Rego}
\IEEEauthorblockA{jotarego@gmail.com}
\and
\IEEEauthorblockN{Allan de M. Martins}
\IEEEauthorblockA{allan@dee.ufrn.br}
}

\maketitle

\begin{abstract}

Recent studies have demonstrated that correntropy is an efficient tool for analyzing higher-order statistical moments in nonGaussian noise environments. Although it has been used with complex data, some adaptations were then necessary without deriving a generic form so that similarities between complex random variables can be aggregated. This paper presents a novel probabilistic interpretation for correntropy using complex-valued data called complex correntropy. An analytical recursive solution for the maximum complex correntropy criterion (MCCC) is introduced as based on the fixed-point solution. This technique is applied to a simple system identification case study, as the results demonstrate prominent advantages regarding the proposed cost function if compared to the complex recursive least squares (RLS) algorithm. By using such probabilistic interpretation, correntropy can be applied to solve several problems involving complex data in a more straightforward way.

\end{abstract}

\begin{IEEEkeywords}
complex-valued data correntropy, maximum complex correntropy criterion, fixed-point algorithm.

\end{IEEEkeywords}

\IEEEpeerreviewmaketitle

\section{Introduction}

Defining the relationship between the input and output signals in a given system is a common problem widely found in distinct engineering areas \cite{exemplo1,exemplo2,exemplo3,voz}. The classic regression solution is then extensively adopted using the mean square error (MSE) as a cost function in order to minimize the error between the input signal and the desired output. However, many authors have developed methods based on correntropy as a cost function in the last few years since such approach improves the fitting performance in nonGaussian noise environments \cite{propriedadesCorrentropia,corretropia2,Fontes2015}.

Correntropy is a similarity measure between two variables, which contains information from all the even statistical moments, being a generalization of the correlation concept \cite{Santamaria2006a}. Several techniques have proposed the use of correntropy in adaptive system training, thus demonstrating its excellent performance in practical applications such as noise cancellation in speech signals \cite{xu2008pitch}, system identification  \cite{linhares2015fuzzy,liu2007correntropy,liu2013correntropy,rego2linear}, and face recognition \cite{he2011maximum}, where the errors are typically nonGaussian.

On the other hand, many cases typically involve a processed signal that belongs to the complex domain, as the complex RLS algorithm is widely used as a possible solution \cite{livroRLSComplexo}. However, few studies have explored the use of correntropy as a cost function in problems involving complex-valued data. The work developed in \cite{koreano} presents a complex-valued blind equalization algorithm for quadrature amplitude modulation (QAM) and complex channel environments based on the correntropy criterion. The study is motivated by the improved performance achieved by information theoretic learning (ITL) methods when compared to MSE-based approaches. A robust adaptive carrier frequency offset (CFO) algorithm was introduced in \cite{iraniano} for orthogonal frequency division multiplexing (OFDM) purposes, which also deals with QAM and phase-shift keying (PSK) complex symbols. However, correntropy as applied to complex-valued data has not yet been properly formalized.

Within this context, this paper presents a new probabilistic interpretation for correntropy with complex-valued data, which is defined as complex correntropy. It is based on the probability function in multidimensional spaces using the Parzen estimator. In addition, a novel analytical recursive solution to the MCCC applied as a cost function based on the fixed-point solution is proposed. The results demonstrate the advantages of the proposed cost function in nonGaussian environments when compared to the RLS algorithm for noise cancellation purposes. The remaining sections of this paper are organized as follows. Section II reviews the probabilistic interpretation of correntropy and extends this concept to complex-valued data. Section III presents a closed form recursive solution to MCCC. Simulation results are presented in section IV, while a proper comparative analysis with the RLS algorithm performance is presented. Finally, relevant conclusions are given in Section V.

\section{ Probabilistic Interpretation of Correntropy }
\label{PI}

This section reviews the probabilistic interpretation of correntropy applied to real-valued data, so that it can be further extended to complex-valued data.

\subsection{Correntropy Applied to Real-Valued Data}
\label{CRN}

Correntropy is directly related to the probability regarding how similar two random variables are. In fact, correntropy is the exact estimate of such probability when a Parzen estimator is used for the joint probability \cite{livroitl}. Firstly, let us consider two arbitrary scalar random variables $X$ and $Y$, as the correntropy between then is defined as:

\begin{equation}\label{c1}
V(X,Y) = \hat{P}(X = Y) = \int_{-\infty}^{\infty}  \int_{-\infty}^{\infty} \hat{f}_{XY}(x,y) \delta(x-y) \mathrm{d}x\mathrm{d}y
\end{equation}

In most cases, the real distribution is unknown and only a finite number of data samples ${(x_{n},y_{n}), n= 1,2,...N}$ is available. However, it is possible to use the L-dimensional Parzen estimation with a Gaussian kernel to obtain $f_{XY}(x,y)$ as:

\begin{equation}\label{parzenL}
\hat{f}_{X_{1},X_{2},...X_{L}}(x^{1},x^{2},...,x^{L}) = \frac{1}{N}\sum\limits_{n=1}^N \prod \limits_{l=1}^L  G_{\sigma}(x^{l}-x^{l}_{n})
\end{equation}
where $G_{\sigma}(x)$ is defined as

\begin{equation}
G_{\sigma}(x) = \frac{1}{\sqrt{2\pi}\sigma}exp \left ( -\frac{x^2}{2\sigma^2} \right )
\end{equation}

Notation $x^{l}_{n}$ represents the $n$-th data sample for the $l$-th component of the L-dimensional random vector while $\sigma$ is the kernel bandwidth, also known as the kernel size. In order to define correntropy for the real domain, the work presented in \cite{integralxy} considers L=2 in equation (\ref{parzenL}):

\begin{equation}\label{Parzenclassica}
\hat{f}_{XY}(x,y)= \frac{1}{N} \sum\limits_{n=1}^N  G_{\sigma}( (x - x_{n}) + (y - y_{n}))
\end{equation}

Substituting (\ref{Parzenclassica}) in (\ref{c1}) gives:

\begin{equation}\label{c4}
V(x,y) = \int \int  \frac{1}{N} \sum\limits_{n=1}^N  G_{\sigma}( (x - x_{n}) + (y - y_{n})) \delta(x-y) \mathrm{d}x\mathrm{d}y
\end{equation}

If $x=y$, equation (\ref{c4}) can be rewritten as:

\begin{equation}\nonumber
V(X,Y) = \int_{-\infty}^{\infty}  \frac{1}{N} \sum\limits_{n=1}^N G_{\sigma}( (x - x_{n}) + (y - y_{n})) \mathrm{d}u \Big|_{x=y=u} 
\end{equation}

\begin{equation}\label{c2}
V(X,Y) = \int_{-\infty}^{\infty}  \frac{1}{N} \sum\limits_{n=1}^N  G_{\sigma}( (u - x_{n}))  G_{\sigma}(u - y_{n})) \mathrm{d}u 
\end{equation}
where $u$ represents the value assumed by $x$ and $y$ over the line $x=y$.

Equation (\ref{c2}) can be solved as

\begin{equation}\label{classica}
V(X,Y) = \frac{1}{N}\sum\limits_{n=1}^N G_{\sigma}(x_{n}-y_{n})
\end{equation}
which corresponds to the expression that represents correntropy when applied to real-valued random variables $X$ and $Y$ \cite{livroitl}.

\subsection{Correntropy Applied to Complex-Valued Data}
\label{CCN}

Statistical signal processing in the complex domain has traditionally been viewed as a straightforward extension of the corresponding algorithms in the real domain~\cite{koreano}. This paper is then supposed to present a probabilistic interpretation based on Parzen estimator defined according to equation (2) to measure the similarity between two complex variables. Assuming two random complex variables $C_{1}=X+j\,Z$ and  $C_{2} = Y+j\,S$, where $C_{1},C_{2} \in \mathbb{C}$, and $X,Y,Z,S$  are real-valued random variables, it is possible to use correntropy to measure the probability for which such complex numbers are equal. For this purpose, the correntropy concept must be extended for more than two variables, what can be performed when assuming that the probability regarding $C_{1} = C_{2}$ causes the respective real and imaginary parts of $C_{1}$ and $C_{2}$ to be the same. Such probability can be stated in the form:

\begin{equation}
P(C_{1} = C_{2}) = P(X = Y \text{ and } Z = S)
\end{equation}

By using correntropy, such probability can be estimated as:

\begin{equation}
V(C_{1},C_{2})= \! \hat{P}(C_{1} = C_{2}) 
\end{equation}

The probability interpretation of correntropy can be used to estimate the joint probability density $f_{XYZS}$ as:

\begin{equation}\label{integral4}
\begin{split}
V(C_{1},C_{2}) &= \\
 &= \int_{-\infty}^{\infty} \int_{-\infty}^{\infty} \int_{-\infty}^{\infty} \int_{-\infty}^{\infty} \hat{f}_{XYZS}(x,y,z,s) \delta(x-y) \delta(z-s) \mathrm{d}x\mathrm{d}y \mathrm{d}z\mathrm{d}s
\end{split}
\end{equation}
If $x=y$ and $z=s$, equation (\ref{integral4}) can be rewritten as:

\begin{equation}\nonumber
V(C_{1},C_{2})=\int_{-\infty}^{\infty} \int_{-\infty}^{\infty}  \! \hat{f}_{XYZS}(x,y,z,s) \, \mathrm{d}u_{1}\mathrm{d}u_{2} \Big|_{x=y=u_{1}, z=s=u_{2}} 
\end{equation}

\begin{equation}\label{integraldupla}
V(C_{1},C_{2})=\int_{-\infty}^{\infty} \int_{-\infty}^{\infty}  \! \hat{f}_{XYZS}(u_{1},u_{1},u_{2},u_{2}) \, \mathrm{d}u_{1}\mathrm{d}u_{2} 
\end{equation}

It is then possible to replace $\hat{f}_{XYZS}$ for the Parzen estimator defined in equation (\ref{parzenL}) using  $L = 4$:

\begin{equation}\nonumber
\begin{split}
=\int_{-\infty}^{\infty} \int_{-\infty}^{\infty} \frac{1}{N}\sum\limits_{n=1}^N  G_{\sigma}(x-x_{n}) \cdot G_{\sigma}(y-y_{n}) + \\
G_{\sigma}(z-z_{n}) \cdot G_{\sigma}(s-s_{n})  \mathrm{d}u_{1} \mathrm{d}u_{2} \Big|_{x=y=u_{1},z=s=u_{2}}
\end{split}
\end{equation}

\begin{equation}\nonumber
\begin{split}
=\int_{-\infty}^{\infty} \int_{-\infty}^{\infty} \frac{1}{N}\sum\limits_{n=1}^N  G_{\sigma}(u_{1}-x_{n}) \, G_{\sigma}(u_{1}-y_{n}) + \\
G_{\sigma}(u_{2}-z_{n}) \, G_{\sigma}(u_{2}-s_{n})  \mathrm{d}u_{1} \mathrm{d}u_{2} 
\end{split}
\end{equation}

\begin{equation}\label{dupla}
\begin{split}
=\frac{1}{N}\sum\limits_{n=1}^N  \int_{-\infty}^{\infty} \int_{-\infty}^{\infty} G_{\sigma}(u_{1}-x_{n}) \, G_{\sigma}(u_{1}-y_{n}) + \\
G_{\sigma}(u_{2}-z_{n}) \, G_{\sigma}(u_{2}-s_{n})  \mathrm{d}u_{1} \mathrm{d}u_{2} 
\end{split}
\end{equation}

Solving the double integral in (\ref{dupla}) gives:

\begin{equation}\label{final}
V(C_{1},C_{2})=\frac{1}{N}\sum\limits_{n=1}^N  G_{\sigma \sqrt{2}}(x_{n} - y_{n}) \,G_{\sigma \sqrt{2} }(z_{n} - s_{n}) 
\end{equation}

Equation (\ref{final}) is then defined as correntropy for two complex random variables or simply complex correntropy. There are no assumptions or restrictions for its application to generic data e.g. constant modulus or argument, since  it represents a complete measure of similarity between two random variables. It is important to understand the effect of the estimator when computing correntropy as a probability estimation, considering that the case where the imaginary part is equal to zero in both random variables could be misinterpreted. As a result of the Parzen estimator effect, it does not lead to correntropy as defined in equation (\ref{classica}), although the same result can be obtained by only adjusting the kernel size. This is the reason why real-valued correntropy does not add up to 1 if $x=y$. Equation (\ref{final}) can also be further analyzed according to its respective Taylor series expansion. In addition, it is possible to write the average sum as the expected value in the Parzen estimator, which leads to:

\begin{equation}\nonumber
V(C_{1},C_{2}) = \frac{1}{2 \pi \sigma^2} \sum\limits_{m=0}^\infty  \dfrac{(-1)^m}{2^m \sigma^{2m} n! }E[(x-y)^{2m} + (z-s)^{2m}]
\end{equation}

\begin{equation}\label{taylor}
V(C_{1},C_{2}) = \frac{1}{2 \pi \sigma^2} + \frac{k_{1}}{\sigma^4}E[(C_{1},C_{2})(C_{1},C_{2})^{*} + h_{\sigma}(C_{1},C_{2}) 
\end{equation}
considering that 
\begin{equation}
h_{\sigma}(C_{1},C_{2}) = \frac{1}{2 \pi \sigma^2}  \sum\limits_{m=2}^\infty  \dfrac{(-1)^m}{2^m \sigma^{2m} n! }(E[(x-y)]^{2m} + E[(z-s)^{2m}])
\end{equation}
where $h_{\sigma}(C_{1},C_{2})$  is a term that contains all higher-order moments, whose components in the denominator depend on $\sigma$ considering that the first term includes $\sigma^{6}$. 

According to equation (\ref{taylor}), the higher-order terms represented by $h_{\sigma}$ tend to zero faster than the second term as $\sigma$ increases. It is worth to mention that the second term corresponds exactly to the covariance involving two complex variables $C_{1}$ and $C_{2}$. Hence, as the kernel size increases, the complex correntropy tends to the covariance analogously to the conventional one.

\section{Maximum Complex Correntropy Criterion}
\label{FPS}

% \begin{figure}
% \centering
% 	\includegraphics[width=3.2in]{Arquitetura.jpg}
% 	\caption{Typical system identification}
% \label{figuraSistema}
% \end{figure}

\begin{figure}[h!]
\centering
\input{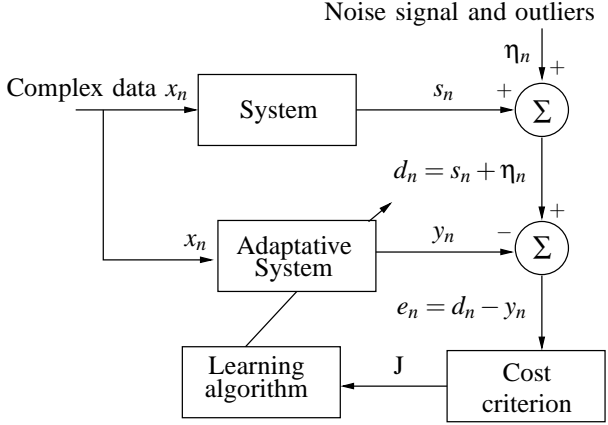}
\caption{Typical system identification}
\label{figuraSistema}
\end{figure}

A typical system identification task is represented in Fig. \ref{figuraSistema}. Since correntropy has been previously defined in the complex domain, it is necessary to establish the MCCC. Let the new cost function $J_{MCCC}$ be the maximum complex correntropy between two random complex variables, where $D$ is the desired signal and $Y$ is the filter output, while $D, Y$ are complex-valued random variables.  

\begin{equation} \label{CCJ}
J_{MCCC} = \underset{w} max \, V(D,Y) 
\end{equation}

The filter output consists in the combination of the system input~$x$ and weights $w$, where $x,w \in \mathbb{C}$. Then, it gives:

\begin{equation}
\begin{split}
Y = w\,X \\
w= w^{re}+j\,w^{im}
\end{split}
\end{equation}
and

\begin{equation}
\begin{split}
D = D^{re} + j\,D^{im} \\
X = X^{re} + j\,X^{im}
\end{split}
\end{equation}

In order to use correntropy in this case, let us assume $C_{1} = D$ and $C_{2} = Y$ in equation (\ref{final}).Then $x_{n}$ becomes $d^{re}_{n}$, $y_{n}$ corresponds to $w^{re}_{n}\,x^{re}_{n} - w^{im}_{n}\,x^{im}_{n}$, and so on. The correntropy-based cost function can be expressed as:

\begin{equation}\label{CCJ2}
\begin{split}
J_{MCCC} = \frac{1}{N}\sum\limits_{n=1}^N  G_{\sigma \sqrt{2}}(d^{re}_{n} - (w^{re}\,x^{re}_{n} - w^{im}\,x^{im}_{n})) \\
G_{\sigma \sqrt{2}}(d^{im}_{n} - (w^{re}\,x^{im}_{n} + w^{im}\,x^{re}_{n}  )) )
\end{split}
\end{equation}

The fixed-point solution for the optimal weights can be obtained by setting the cost function derivative to zero in equation (\ref{CCJ2}):

\begin{equation}\label{derizeroa1ea2}
\dfrac{\partial J_{MCCC}}{\partial w^{re}} = 0 \quad \text{and} \quad \dfrac{\partial J_{MCCC}}{\partial w^{im}} = 0
\end{equation}
isolating $w^{re}$ and $w^{im}$ gives:

\begin{equation}\label{a1}
w^{re} = \dfrac{\sum\limits_{n=1}^N g_{1}(d,w,x) \, g_{2}(d,w,x)  (d^{re}_{n}\,x^{re}_{n} + d^{im}_{n}\,x^{im}_{n} ) }{\sum\limits_{i=n}^N g_{1}(d,w,x) \, g_{2}(d,w,x)((x^{re}_{n})^2 + (x^{im}_{n})^2 )}
\end{equation}
 
\begin{equation}\label{a2}
w^{im} = \dfrac{\sum\limits_{n=1}^N g_{1}(d,w,x) \, g_{2}(d,w,x)  (d^{im}_{n}\,x^{re}_{n} - d^{re}_{n}\,x^{im}_{n} )}{\sum\limits_{n=1}^N g_{1}(d,w,x) \, g_{2}(d,w,x)((x^{re}_{n})^2 + (x^{im}_{n})^2 )}
\end{equation}
where

\begin{equation}\label{pf1}
g_{1}(d,w,x) = G_{\sigma \sqrt{2}}(d^{re}_{n} - (w^{re}\,x^{re}_{n} - w^{im}\,x^{im}_{n})) 
\end{equation}

\begin{equation}\label{pf2}
g_{2}(d,w,x) = G_{\sigma \sqrt{2}}(d^{im}_{n} - (w^{re}\,x^{im}_{n} + w^{im}\,x^{re}_{n}))
\end{equation}

The result provided by equations (\ref{pf1}) and (\ref{pf2}) represents the iterative solution for $w^{re}$  and $w^{im}$. Even though convergence is achieved after a few iterations, each one of them requires the computation of the whole sum, which is inadequate to real-time learning. A fixed-point stochastic recursive solution can then be derived as inspired by \cite{singh2010closed} and based on equations (23) and (24). Firstly, let us define the iterations for (\ref{pf1}) and (\ref{pf2}) as:

\begin{equation}
w^{re}[n] = \frac{P[n]}{R[n]}
\end{equation}

\begin{equation}
w^{im}[n] = \frac{Q[n]}{R[n]}
\end{equation}

Applying a stochastic approach to the sum gives:

\begin{equation}\label{pe}
P[n] = P[n-1] + g_{1}(d,w,x) \, g_{2}(d,w,x)  (d^{re}_{n-1}\,x^{re}_{n-1} + d^{im}_{n-1}\,x^{im}_{n-1} )
\end{equation}

\begin{equation}\label{que}
Q[n] = Q[n-1] + g_{1}(d,w,x) \, g_{2}(d,w,x)  (d^{im}_{n-1}\,x^{re}_{n-1} - d^{re}_{n-1}\,x^{im}_{n-1} )
\end{equation}

\begin{equation}\label{erre}
R[n] = R[n-1] + g_{1}(d,w,x) \, g_{2}(d,w,x)((x^{re}_{n-1})^2 + (x^{im}_{n-1})^2 )
\end{equation}

In order to implement the recursive expressions represented by (\ref{pe}), (\ref{que}), and (\ref{erre}), it is necessary to compute the initial parameters $w^{re}$ and $w^{im}$, as well as compute the initial values for $P[0],Q[0],R[0]$.

\section{Simulation and Results}

In order to evaluate the MCCC performance, the complex RLS algorithm presented in \cite{livroRLSComplexo} has been adopted for comparison purposes. Besides, the weight signal-to-noise ratio (WSRN) is also considered in the analysis of results as in \cite{singh2010closed}, since it quantifies convergence and misadjustment rates properly in decibels as:

\begin{equation}
WSNR_{db} = 10 \log_{10} \left (  \frac{ \bar{w}\,\bar{w}^{*}  }{ (\bar{w} - w[n])(\bar{w} - w[n])^{*} }  \right )
\end{equation}
Where $\bar{w} = \bar{w}^{re} + j\,\bar{w}^{im}$ is the proper weight chosen for the simulation tests and $w[n] = w^{re}[n] + j\,w^{im}[n]$ is the weight computed by the aforementioned methods in the $n$-th iteration.

\begin{figure}[h!]
\centering
	\includegraphics[width=3.5in]{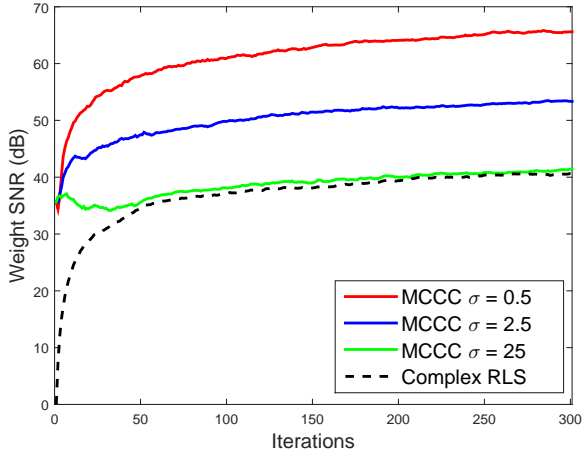}
	\caption{Weight SNR plots for MCC fixed point  } 
\label{wsnr}
\end{figure}

The desired signal is contaminated with nonGaussian noise whose PDF (probability density function) is $0.95\mathcal{N}(0.0,0.05) + 0.05\mathcal{N}(0.0,5.0)$, where $\mathcal{N}(\mu,\sigma)$ is a normal Gaussian distribution with mean $\mu$ and variance $\sigma^2$. The authors in \cite{singh2010closed} also employ the aforementioned PDF to represent the noise and evaluate robustness of the fixed-point MCC algorithm compared with its RLS counterpart, although data only comprises the real domain. The noise signal $\eta_{n}$ is then generated in this work, where $\eta \in \mathbb{C}$ and  $\eta_{n} = \eta^{re}_{n} + j\,\eta^{im}_{n}$, while $\eta^{re}_{n}$ and $\eta^{im}_{n} \in \mathbb{R}$ and follow the described PDF.

After 300 iterations, the results shown in Fig. \ref{wsnr} could be obtained. The curves represent the average when using 50 Monte Carlo trials, as the weights always start from random values. Fast convergence is achieved by both methods as in \cite{singh2010closed}, but improved performance is achieved when the kernel size is =0.5. It can be stated that the proposed approach is able to ignore outliers. The kernel size in equation (\ref{taylor}) behaves as a parameter that weights both second-order $(m=1)$ and higher-order moments. As $\sigma$ becomes higher than unity, the high-order moments decrease faster as the achieved results are closer to the ones provided by the conventional complex RLS solution.

\section{Conclusions}

This paper has presented the extension of the correntropy concept to complex-valued data in an approach defined as complex correntropy. A significant contribution of this work lies in obtaining the expression for the complex correntropy from its respective probabilistic interpretation. Besides, a recursive algorithm based on fixed-point solution has been introduced, which can be used to derive the MCCC. Simulation tests have also demonstrated that the proposed method presents high convergence rates, but with higher efficiency when dealing with outlier environments if compared to the complex RLS approach. It is then reasonable to state that correntropy can now be applied to the solution of distinct problems involving complex data in a more straightforward way.

\ifCLASSOPTIONcaptionsoff
  \newpage
\fi



\bibliographystyle{IEEEtran}
\bibliography{DSP}
%\begin{thebibliography}{1}
%\end{thebibliography}



%\vfill


\end{document}